# Entanglement measures in quantum and classical chaos

Arul Lakshminarayan, Jayendra Bandyopadhyay, M. S. Santhanam, and V. B. Sheorey

*Abstract*—Entanglement is a Hilbert-space based measure of nonseparability of states that leads to unique quantum possibilities such as teleportation. It has been at the center of intense activity in the area of quantum information theory and computation. In this paper we discuss the implications of quantum chaos on entanglement, showing how chaos can lead to large entanglement that is universal and describable using random matrix theory. We also indicate how this measure can be used in the Hilbert space formulation of classical mechanics. This leads us to consider purely Hilbert-space based measures of classical chaos, rather than the usual phase-space based ones such as the Lyapunov exponents, and can possibly lead to understanding of partial differential equations or nonintegrable classical field theories.

## I. INTRODUCTION

QUANTUM entanglement was first discussed by Shrödinger [1] to point out the "weirdness", or non-classicality, implied by the laws of quantum mechanics. It is the property that leaves the cat, named after him and inhabiting a closed box with a deadly trigger, in the Vishwamitra space between the living and the dead. It is however a natural consequence of the *Hilbert* space of the states of the cat. These states get *entangled* with the states of the trigger device inside the box, and the consequence is the much discussed limbo.

In a more prosaic setting, consider two spin-half particles say labelled by $A$ and $B$. Use the standard $S_z$ basis $\{|0\rangle, |1\rangle\}$ ("up","down") for each particle. Then the spin singlet state:

$$|\psi_{AB}\rangle = \frac{1}{\sqrt{2}}(|0\rangle_A|1\rangle_B - |1\rangle_A|0\rangle_B)$$

is an example of an entangled state. This two-particle state forbids the allocation of a pure state to either of the particles, it implies a correlation between the two that is over and above that which is classically possible, it violates the Bell-type inequalities maximally [2].

Recall that if the particle $B$ were in a pure state, it should be possible to orient a Stern-Gerlach apparatus in such a manner that provided many copies of such pairs of particles, we can be sure of the consequences of the spin measurements ("up" or "down") every time. For the state $|\psi_{AB}\rangle$ if measurement of particle $A$ is done for the $S_z$ component, the $B$ component is known, but only to those that measured $A$, and who have prior knowledge of the two-particle state. It leaves the two-particle states in either $|0\rangle_A|1\rangle_B$ or $|1\rangle_A|0\rangle_B$ with equal probability. Hence spin measurement on $B$ can never be deterministic, no matter which way we orient the Stern-Gerlach apparatus. Particle B will behave exactly like a classical coin, realizing "up" or "down" randomly. In fact particle $A$ can be measured along *any* direction and the result is the same. In other words, the state of particle $B$ cannot be written as a pure state, it is described by an impure state, and specified by a density matrix. This follows from the fact that there exist no states that belong solely to the individual Hilbert space of particles $A$ and $B$ such that their outer (tensor) product is the spin singlet state $|\psi_{AB}\rangle$ [2].

## II. QUANTIFYING ENTANGLEMENT

How do we quantify entanglement? Is it observable? These are two different questions, the latter is still unclear, although there are proposals for indirect measurements [4]. Quantifying entanglement is also not easy or clear in the case when more than two subsystems are involved. (The subsystem can consist of one or more particles.) Thus while there are many new advances in the theory of entanglement it is by no means complete. The simplest case, with which we will concern ourselves below, is when we have a system in a pure quantum state, consisting of two subsystems ("bipartite") among which we study the entanglement. In this case the reduced density matrices of the subsystems, carry information about entanglement. If there is no entanglement, then the reduced density matrices will correspond to density matrices of pure states and hence the von Neumann entropy of the reduced density matrices (RDMs) is a measure of entanglement, it will vanish for unentangled states.

Suppose that the state space of a bipartite quantum system is $\mathcal{H} = \mathcal{H}_1 \otimes \mathcal{H}_2$, where $\dim \mathcal{H}_1 = N \le \dim \mathcal{H}_2 = M$, and $\dim \mathcal{H} = d = NM$. If $\rho = \sum_i p_i |\phi_i\rangle\langle\phi_i|$ is an ensemble representation of an arbitrary state in $\mathcal{H}$, the entanglement (of formation) is found by minimizing $\sum_i p_i E(|\phi_i\rangle)$ over all possible ensemble realizations. Here $E$ is the von Neumann entropy of the RDMs of the state $|\phi_i\rangle$ belonging to the ensemble, *i.e.*, its entanglement. For pure states $|\psi\rangle$ there is only one unique term in the ensemble representation and the entanglement is simply the von Neumann entropy of the RDM.

The two RDMs of the bipartite state $|\psi\rangle$ are $\rho_1 = \text{Tr}_2(|\psi\rangle\langle\psi|)$ and $\rho_2 = \text{Tr}_1(|\psi\rangle\langle\psi|)$. The Schmidt decomposition of $|\psi\rangle$ is the optimal representation in terms of a product basis and is given by

$$|\psi\rangle = \sum_{i=1}^{N} \sqrt{\lambda_i} |\phi_i^{(1)}\rangle |\phi_i^{(2)}\rangle, \qquad (1)$$

where $0 < \lambda_i \le 1$ are the (nonzero) eigenvalues of either RDMs and the vectors are the corresponding eigenvectors. The von Neumann entropy $S_V$ is the entanglement $E(|\psi\rangle)$ given by

$$S_V = -\text{Tr}_l(\rho_l \ln \rho_l) = -\sum_{i=1}^{N} \lambda_i \ln(\lambda_i), \; ; \; l = 1, 2. \qquad (2)$$

There is only one entropy as the two RDMs are unitarily related, they share the same nonzero eigenvalues. Given the constraint that $\rho_i$ are density matrices, the maximum value that $S_V$ can take is $\log(N)$. We note that the Schmidt decomposition is identical to the Singular Value Decomposition of Linear Algebra

Arul Lakshminarayan is with the Department of Physics, Indian Institute of Technology Madras, (corresponding author, e-mail: arul@iitm.ac.in), Jayendra Bandyopadhyay and V. B. Sheorey are with the Physical Research Laboratory, Ahmedabad, M. S. Santhanam is at the Max Planck Institute for Complex Systems, Dresden, Germany.

and does not apply if more than two Hilbert spaces are involved. While it may appear odd that a general state in an $NM$ dimensional Hilbert space is represented as a sum over only $N$ terms, it should be kept in mind that the basis, $|\phi_i^{(1)}\rangle|\phi_i^{(2)}\rangle$ are themselves state dependent, it is just that the basis is still of the form of an outer product. Thus for bipartite pure states entanglement may be equated to separability.

### III. CHAOS AND QUANTUM ENTANGLEMENT

While entanglement has been studied as a "purely quantum" phenomena, a similar statement is often made of chaos, as a purely "classical phenomena", which either means that there is no true chaos, as the world is quantum mechanical, or that something pretty strange is happening in the quantum-classical transition that as radical a phenomenon as chaos miraculously appears, and something as unusual as entanglement disappears. It is by now well established that classical chaos leaves unique "fingerprints" on quantization [3]. Notable, is the appearance of classically unstable periodic orbits in quantum wavefunctions ("scarring") and the applicability of ensemble theories like the random matrix model to isolated simple few-degree of freedom systems, such as two quartic oscillators, or the hydrogen atom in a magnetic field.

Several research groups have started exploring the connections between chaos and entanglement [5], [7], using systems such as the standard map, the kicked top, coupled quartic oscillators, spin chains the Jaynes-Cummings Hamiltonian and quantum dots [4]. Some analytical aspects of this work are due to modelling using random matrix theory (RMT) [3]. Semiclassical approaches are very difficult and are beginning to be studied. Generically we will consider Hamiltonians of the form $H = H_1 + H_2 + \alpha H_{12}$, where $H_1$ and $H_2$ are the Hamiltonians of the two subsystems that interact through the third term. We may consider the cases when $H_i$ are not chaotic (integrable), but $H$ could be chaotic (nonintegrable), or when both $H_i$ and $H$ are nonintegrable. In fact entanglement and quantum information theory in general has been conjectured to provide a basis for studying complex quantum systems [6]. It appears that entanglement may be strange, but it is not uncommon.

#### A. Top Models

As the Hilbert space dimension and chaos have roles to play in entanglement, coupled large spins are attractive models. Rather than specifying Hamiltonians for continuous time evolution, we specify unitary operators for finite, discrete time translation, this may be thought of as Floquet operators of kicked Hamiltonians, or quantum maps [3] The unitary time evolution operator, $U_T$, considered here is [7]:

$$U_T = (U_1 \otimes U_2) U_{12}^\epsilon = [(U_1^f U_1^k) \otimes (U_2^f U_2^k)] U_{12}^\epsilon \quad (3)$$

$$\text{where} \quad U_l^f = e^{-i\pi J_{y_l}/2}, U_l^k = e^{-ik(J_{z_l}+\alpha_l)^2/2j_l}, \quad (4)$$

$$\text{and} \quad U_{12}^\epsilon = e^{-i\epsilon J_{z_1} J_{z_2}/\sqrt{j_1 j_2}}. \quad l = 1, 2.$$

The $U^f$ operators are simply rotation of each top about the $y$ axis by $\pi/2$, while the others are due to periodic $\delta$-function kicks. The $U^k$ operators are torsion about $z$-axis, and are parametrized by the strength $k$ that controls the amount of classical chaos in the classical models, and the final term describes the spin-spin coupling. When either of the constants, $\alpha_1$ or $\alpha_2$, is not zero the parity symmetry $RH(t)R^{-1} = H(t)$, where $R = \exp(i\pi J_{y_1}) \otimes \exp(i\pi J_{y_2})$, is broken. The dimensionality of the Hilbert spaces are $N = 2j_1 + 1$ and $M = 2j_2 + 1$.

We note that for the parameter values considered below, the nearest neighbor spacing distribution (NNSD) of the eigenangles of $U_T$ is Wigner distributed, which is typical of quantized chaotic systems with time reversal symmetry [3]. Entanglement production of time evolving states under $U_T$ have been studied for two different initial states. (1) The initial state is a product of directed angular momentum states, placed in the chaotic sea of phase space. This is a completely unentangled state. (2) The initial state is maximally entangled and is given by: $\langle m_1, m_2|\psi(0)\rangle = \delta_{m_1 m_2}/N$.

These initial states are evolved under $U_T$, and the results are displayed in Fig. 1. In the first case, initially both the von Neumann entropy and the linearized entropy ($S_R = 1 - \text{Tr}_1(\rho_1^2)$), are zero, but with time evolution both entropies start increasing

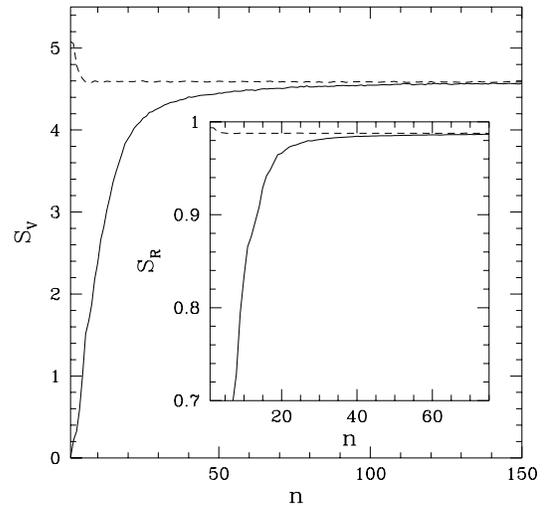

Fig. 1. Entanglement saturation of a completely unentangled initial state (solid line) and a maximally entangled initial state (dotted line) under time evolution operator $U_T$. Here $k = 3, \epsilon = 0.1$ and the phases $\alpha_1 = \alpha_2 = 0.47$. Inset shows similar behaviour of linear entropy.

and saturate, apart from small fluctuations, at values *less* than the maximum possible ($\log(N)$.)

In fact we have pointed out [7] the *distribution* of the eigenvalues of the RDM, the $\lambda_i$, as this must follow from the fact that eigenstates are described by RMT. The two RDMs have the structure $A^\dagger A$ and $AA^\dagger$, $A$ containing the vector components of the bipartite system. Thus using RMT implies that $A$ has Gaussian distributed random entries, and the distribution of the eigenvalues of $A^\dagger A$ are then already known, and can be directly used to estimate entanglement.

The density of the eigenvalues of the RDM $\rho_1$ is given by

$$\begin{aligned} f(\lambda) &= \frac{NQ}{2\pi} \frac{\sqrt{(\lambda_{max}-\lambda)(\lambda-\lambda_{min})}}{\lambda} \\ \lambda_{min}^{max} &= \frac{1}{N}\left(1 + \frac{1}{Q} \pm \frac{2}{\sqrt{Q}}\right), \end{aligned} \quad (5)$$

where $\lambda \in [\lambda_{min}, \lambda_{max}]$, $Q = M/N$ and $Nf(\lambda)d\lambda$ is the num-

ber of eigenvalues within $\lambda$ to $\lambda + d\lambda$. This has been derived under the assumption that both $M$ and $N$ are large. Note that this predicts a range of eigenvalues for the RDMs that are of the order of $1/N$. For $Q \neq 1$, the eigenvalues of the RDMs are bounded away from the origin, while for $Q = 1$ there is a divergence at the origin. All of these predictions are seen to be borne out in numerical work with coupled tops.

Fig. 2 shows how well the above formula fits the eigenvalue distribution of reduced density matrices corresponding to the eigenstates of the coupled tops. Time evolving states also have the same distribution. This figure also shows that the probability of getting an eigenvalue outside the range $[\lambda_{min}, \lambda_{max}]$ is indeed very small. From this distribution we can derive expressions for the expected saturation entanglement, which is of the form $S_V = \log(N \gamma(N/M))$, where $\gamma$ is a function of the ratio of the Hilbert space dimensionalities, and as $M \longrightarrow \infty$, it tends to unity. For $M = N$, $\gamma = 1/\sqrt{e}$. For a much more detailed account of the coupled tops and related entanglement issues we refer to recent works [7].

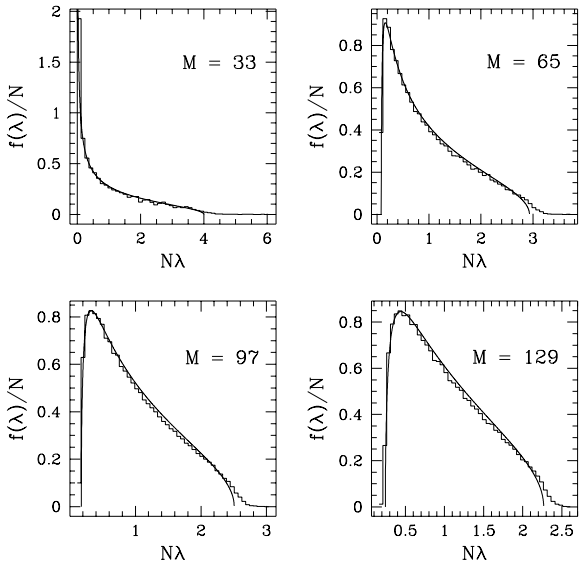

Fig. 2. Distribution of the eigenvalues of the RDMs of coupled kicked tops, averaged over all the eigenstates ($N = 2j_1 + 1 = 33$). Solid curves corresponds to the theoretical distribution function Eq. (5).

### B. Quartic Oscillators

We have also begun the study of entanglement properties of anharmonic oscillator chains [8]. The simplest case considered is a 2D potential $V(q_1, q_2) = (q_1^2 + q_2^2)/2 + \alpha q_1^2 q_2^2/2$. For $\alpha > 0$ the two oscillators interact and in general the eigenstates are entangled, and for even modest values of $\alpha$ the oscillators are classically chaotic. Earlier investigations had shown up a series of "channel localized" eigenstates, those that are dominated by the straight-line periodic orbits, which correspond to 1D motions along either $q_1$ or $q_2$ directions. We have found that these states are also local minima for entanglement, another clear signature of classical mechanics in something that is considered purely quantum mechanical. There are many open questions, including interesting ones regarding large oscillator chains, such as the FPU.

### IV. ENTANGLEMENT AND CLASSICAL CHAOS

While in fact quantum entanglement is purely a quantum feature, we can ask what happens in the classical limit? Here we address a more modest problem, we wish to define a nonseparability measure in classical mechanics, which must naturally have many similarities to quantum entanglement [9]. While the quantum-classical correspondence has often been studied via the (psuedo) phase-space representation of quantum states, we can also complement this by studying classical mechanics in a Hilbert space. The states are functions in phase-space, the dynamics is, like the quantum, a linear one, governed by the Liouville equation. We note in passing that there are known connections between entanglement and thermodynamic entropy and that the classical measure defined below has similarities to an elusive entropy-like quantity.

Now consider the classical mechanics as described in the phase-space $(q_i, p_i)$, $i = 1, 2, \ldots, n$. The flow is generated by a Hamiltonian $H$ and the Frobenius-Perron (F-P) operator, which is the exponentiation of the Liovillian, is defined through:

$$\mathcal{P}(t) = \exp(-t\{\ , H\}); \; f(q_i, p_i, t) = \mathcal{P}(t) f(q_i, p_i, 0). \quad (6)$$

The F-P operator is unitary. We consider a class of functions that is broad enough to include physically relevant ones. For simplicity, and obvious connections to quantum mechanics, we consider the space of $L^2$ functions. Thus we allow for classical "states" possibly complex functions over the phase-space. The differences between quantum and classical stands out at this early juncture: all the quantum states in the Hilbert space are assumed to be physically realizable, classically the relevant states are those that are at least real everywhere.

Consider again a 4D map, or a 2-degree of freedom flow. Expanding any function $f$ in the four-dimensional phase space in terms of some $L^2$ basis functions $g$ and $h$ for each degree of freedom, we may write:

$$f(q_1, p_1, q_2, p_2) = \sum a(m, n; k, l) g_{m,n}(q_1, p_1) h_{k,l}(q_2, p_2), \quad (7)$$

where the $a(;)$ are expansion coefficients. Now the "reduced density matrix" is simply the partial trace:

$$\rho_1(m, n; m', n') = \sum_{k,l} a(m, n; k, l) a^*(m', n'; k, l), \quad (8)$$

and a similar expression holds for $\rho_2$, while the von Neumann entropy of these reduced density matrices is the entropy of classical "entanglement" and is given by: $S_C = -\text{Tr}_1(\rho_1 \log(\rho_1))$. The analogy to the quantum definitions are evident and the existence of a Schmidt decomposition follows. The entropy $S_C$ can be zero or positive, if it is zero there is no entanglement, and the densities are separable into each degree of freedom, while if it is positive this is not possible. The basis independence of the entropy $S_C$ (the choice of $g$ and $h$) follows from the two trace operations, and thus it represents a physically meaningful quantity, with possible relevance to quantum entanglement, at least numerically, if not conceptually. We clarify that we do not mean to imply by this formalism that there is classical entanglement in the same footing as quantum entanglement. The measure we have defined is function-space based and may be extended to partial differential equations wherein there is no immediate phase space, or trajectories to diverge exponentially or otherwise.

*A. Coupled Standard Map Model*

Consider the classical map defined on the four-torus $T^2 \times T^2$:

$$\begin{aligned} q'_i &= q_i + p_i \pmod 1 \\ p'_i &= p_i - \partial V/\partial q'_i \pmod 1, \end{aligned} \quad (9)$$

where $i = 1, 2$ and the potential $V$ is

$$V = -\sum_{i=1}^{2} \frac{K_i}{(2\pi)^2}\cos(2\pi q_i) + \frac{b}{(2\pi)^2}\cos(2\pi q_1)\cos(2\pi q_2) \quad (10)$$

This is a symplectic transformation on the torus $T^4$ and may be derived from a kicked Hamiltonian in the standard manner [3], [9]. The parameter $b$ controls the interaction between the two standard maps, while $K_i$ determine the degree of chaos in the uncoupled limit. The advantage in studying such maps is our ability to write explicitly the relevant unitary operators, both classical and quantal.

The classical unitary F-P operator over one iteration of the map is $\mathcal{P} = \mathcal{P}_1 \otimes \mathcal{P}_2 \; \mathcal{P}_b$, where $\mathcal{P}_i$ is the F-P operator for the standard map on $T^2$ and acts on the "single-particle" Hilbert spaces, while $\mathcal{P}_b$ is the interaction operator on the entire space. We use as a single-particle basis the Fourier decomposition: $g_{m,n} = h_{m,n} = \exp(2\pi i(m q + n p))$. Then the matrix elements of the F-P operators are:

$$\langle m'_i, n'_i | \mathcal{P}_i | m_i, n_i \rangle = J_{m'_i - m_i}(K_i n'_i)\delta_{n_i - m_i, n'_i}, \quad (11)$$

$$\langle m'_1, n'_1; m'_2, n'_2 | \mathcal{P}_b | \langle m_1, n_1; m_2, n_2 \rangle = J_{l+}(b(n_1 + n_2)/2) J_{l-}(b(n_1 - n_2)/2) \delta_{n_1, n'_1}\delta_{n_2, n'_2}, \quad (12)$$

where $l\pm = ((m_1 - m'_1) \pm (m_2 - m'_2))/2$ and the delta functions are Kronecker deltas. The orders of the Bessel J functions are restricted to the integers.

The number of basis states excited increases with time in general including increasingly larger frequencies. This is indicative of the fine structure that is being created in phase-space by the stretching and folding mechanism of chaos. The frequencies involved appears to increase exponentially in time for chaotic systems, while we may expect a polynomial growth for integrable or near-integrable systems.

The quantization of the symplectic transformation in Eq. (9) is a finite unitary matrix on a product Hilbert space of dimensionality $N^2$, and $N = 1/h$, where $h$ is a scaled Planck constant. The classical limit is the large $N$ limit. The quantization is straightforward as there exists a kicked Hamiltonian generating the classical map [9].

We time evolve analogous initial states, distributions localized at the fixed point $(0, 0, 0, 0)$, both quantum and classical, using the respective propagators and calculate the entanglement in each, as defined above. Due to the infinite dimensionality of the classical Hilbert space and the at least polynomially increasing frequency components, the classical calculations in a truncated Hilbert space lead to rapid loss of accuracy. The case of full-fledged chaos is, classically, computationally prohibitive.

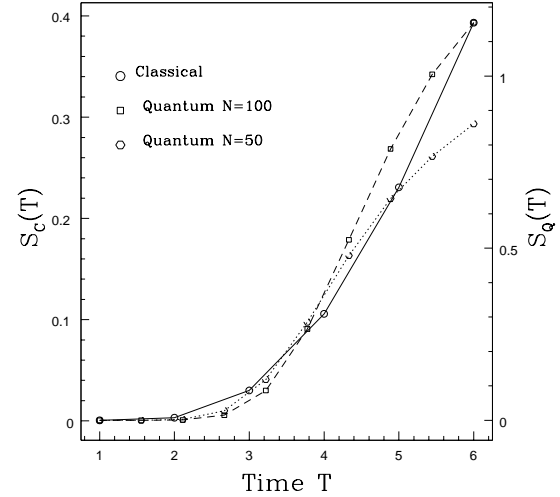

Fig. 3. Classical and quantum entanglement as a function of time for initial states localized at the fixed point at the origin. The interaction $b = 0.05$, while $K_1 = K_2 = 0$.

In Fig. 3 we plot both the entanglements as a function of time for a fixed value of the interaction. We see qualitatively similar behaviour. The quantum entanglement tend to saturate at higher values for larger $N$, in consonance with what we discussed earlier, while the classical entanglement seems to be continuously increasing. This is however difficult to see numerically due to increase of errors, for instance at the highest time shown in the figure, $T = 6$, the normalization deteriorated from unity to about .9948. More elaborate work is needed to establish the properties of the so-called classical entanglement we have introduced here. The larger classical entanglement is to be seen as due to the fine phase-space structures, structures that are not inhibited by the Planck constant.

## REFERENCES

[1] E. Schrödinger, Proc. Camb. Philos. Soc. **31**, 555 (1935).
[2] A. Peres, *Quantum Theory: Concepts and Methods*, (Kluwer Academic Publishers, Dordrecht, 1993); J. Preskill, Lecture Notes at http://www.theory.caltech.edu/people/preskill/ph229; M. A. Nielsen and I. L. Chuang, *Quantum Computation and Quantum Information* ( Cambridge University Press, Cambridge, 2000 ).
[3] F. Haake, *Quantum Signatures of Chaos*, 2nd ed. (Springer-Verlag, Berlin, 2000); L. E. Reichl, *The transition to Chaos in Conservative Classical Systems: Quantum Manifestations*, (Springer-Verlag, New York, 1992).
[4] C. W. J. Beenakker, M. Kindermann, C. M. Marcus, A. Yacoby, cond-mat/0310199.
[5] P. A. Miller and S. Sarkar, Phys. Rev. E **60**, 1542 (1999); R. M. Angelo, K. Furuya, M. C. Nemes, and G. Q. Pellegrino, Phys. Rev. A **64**, 043801 (2001); A. Lakshminarayan, Phys. Rev. E **64**, 036207 (2001);A. Tanaka, H. Fujisaki, and T. Miyadera, Phys.Rev. E **66** 045201(R) 2002); H. Fujisaki, T. Miyadera, and A. Tanaka, Phys. Rev. E **67**, 066201 (2003); S. Bettelli and D. L. Shepelyansky, Phys. Rev. A **67**, 054303 (2003); A. Lahiri and S. Nag, Phys. Lett. **318A**, 6 (2003); A. J. Scott and C. M. Caves, J. Phys. A **36**, 9553 (2003). A. Lakshminarayan, V. Subrahmanyam, Phys. Rev. A **67**, 052304 (2003).
[6] M. A. Nielsen, quant-ph/0208078.
[7] J. N. Bandyopadhyay and A. Lakshminarayanan, Phys. Rev. Lett. **89**, 060402 (2002); Phys. Rev. E. (To Appear, 2003). quant-ph/0307134.
[8] M. S. Santhanam, V. B. Sheorey, A. Lakshminarayan. (Under Preparation).
[9] Arul Lakshminarayan, quant-ph/0107078.(Unpublished).